\documentclass[pre,amsmath,amssymb,showpacs,superscriptaddress]{revtex4}
\usepackage{graphics}
\usepackage{graphicx}
\usepackage{subfigure}
\usepackage{color}
\usepackage{bm}
\usepackage{natbib}

\def\x{{\bm x}}

\def\c{{\bm c}}

\def\u{{\bm u}}

\def\I{{\bm I}}

\def\F{{\bm F}}

\begin{document}
\title{Well-balanced lattice Boltzmann equation for two-phase flows}
  \author{Zhaoli Guo}
 \email[Email:]{zlguo@mail.hust.edu.cn}
 \affiliation{State Key Laboratory of Coal Combustion, Huazhong University of Science and Technology, Wuhan 430074, China}

\begin{abstract}
The standard lattice Boltzmann equation (LBE) method usually fails to capture the physical equilibrium state of a two-phase fluid system, i.e., zero velocity and constant chemical potential. Consequently, spurious velocities and inconsistent thermodynamic density properties are frequently encountered in LBE simulations. In this work, based on a rigorous analysis of the discrete balance equation of LBE, we identify the structure of the force imbalance due to discretization errors from different parts. Then a well-balanced LBE is proposed which can achieve the discrete equilibrium state. The well-balanced properties of the LBE are confirmed by some numerical tests of a flat interface problem and a droplet system.
\end{abstract}

\maketitle
\section{Introduction}
It is a challenging problem for modelling and simulating multi-phase fluid flows due to the complex interfacial
dynamics involving large space and time scales. Over the past three decades, the lattice Boltzmann equation (LBE) based on kinetic models has been developed as an efficient method for multi-phase flows \cite{ref:SucciBook, ref:GuoBook}. The kinetic nature of LBE makes it straightforward to incorporate the intermolecular interactions into the model, which is
recognized as one of the main advantages of LBE that distinguishes it from other numerical methods. However, the LBE also faces some undesirable features in simulating multiphase systems. One fundamental problem is that it can be quite inaccurate at equilibrium state, such that the numerical results are not compatible with exact physical conditions.

Physically, the chemical potential $\mu$ should be constant and the fluid velocity $\u$ should be zero everywhere for a two-phase system at equilibrium, namely it should maintain the equilibrium state
\begin{equation}
{\cal S}\equiv \{\mu=\mbox{constant}, \u=0\},
\end{equation}
from which other state properties, such as interface structure, bulk phase densities, and pressure distribution, can be deduced. However, there are numerous examples in the literature that the LBE yields nonzero spurious velocity (SV) in the vicinity of phase interface \cite{ref:Cristea,ref:Wagner2003,ref:Lee,ref:Shan,ref:Gong2019,ref:LeeRev} for a stationary droplet immersed in another phase, and produces inaccurate density properties such as coexistence curve and density profiles across interface for an equilibrium flat interface \cite{ref:Zheng2005,ref:Wagner2006,ref:Kupershtokn2009,ref:HuangHBPRE2011,ref:Sun2012,ref:Siebert2014,ref:LuoKH2015,ref:ChenBX2015,ref:ZhengLin2017,ref:HuangPRE2019,ref:Khar2019,ref:WagnerPRE2020,ref:WenPRE2020}.
The appearance of SV and inaccurate density properties suggests that the equilibrium state is not reached by the corresponding LBE models. In other words, at the discrete level the forces from different parts of LBE are NOT well balanced. Particularly, the SV is produced by the non-constant chemical potential which serves as a net driven force. The SV and inaccurate equilibrium density properties may lead to unphysical phenomena and numerical instability in some circumstances, and it is necessary to understand the origin to reduce or remove their effects.

In the past decades, a number of efforts have been attempted to identify the force imbalance in LBE at equilibrium state. Particularly, Guo {\it et al.} made a rigorous analysis the discrete balance equations of a LBE for a flat interface problem \cite{ref:GuoPRE2011}, and a necessary condition for discrete force balance was obtained. It was revealed that the condition cannot be achieved by the standard LBE due to discretization errors in the discretizations of pressure gradient and interaction force. Following this guidance, Lou and Guo developed a Lax-Wendroff type LBE (LW-LBE) which can control the force imbalance with an adjustable Courant-Friedrichs-Levy (CFL) number. However, the force imbalance cannot be removed completely since the CFL number cannot be zero. Furthermore, the time step of the LW-LBE is proportional to the CFL number, such that the computation time can be much increased.

Although the total force imbalance equation was obtained in Ref. \cite{ref:GuoPRE2011}, the detailed discrete errors from different parts are not identified, and the structure of the net force is unclear. The aim of the present study is twofold: (i) to give a further theoretical analysis of the force imbalance property of the standard LBE to identify the detailed errors from different parts, and (ii) to develop a well-balanced LBE that can achieve the equilibrium state.

The rest of the paper is organized as follows. In Sec. II, a theoretical analysis of the balance properties of the standard LBE for a special one-dimensional (1D) flat interface system is presented, focusing on the structure of net force due to discretization errors from different parts. Based on the analysis, a well-balanced LBE is presented and analyzed in Sec. III. Then the capability of capturing equilibrium state of the new LBE is tested in Sec. IV by simulating a 1D flat interface and a two-dimensional (2D) droplet system. Finally a brief summary is presented in Sec. V.

\section{Force imbalance of standard LBE }
For simplicity without loss of generality, we consider the following standard two-dimensional nine-velocity (D2Q9) LBE,
\begin{equation}
\label{eq:D2Q9}
f_i(\x+\c_i\Delta t, t+\Delta t) - f_i (\x, t) = \dfrac{1}{\tau}\left(f_i-f_i^{(eq)}\right) +\Delta t \left(1-\dfrac{1}{2\tau}\right) F_i,
\end{equation}
where $f_i(\x, t)$ is the distribution function for particles moving with velocity $\c_i$ at position $\x$ and time $t$, $\Delta t$ is the time step, and $\c_i$ are discrete velocities defined by $\c_0=(0, 0)$, $\c_1=-\c_3=c(1, 0)$, $\c_2=-\c_4=c(0, 1)$, $\c_5=-\c_7=c(1, 1)$, and $\c_7=-\c_8=c(-1, 1)$, with $c=\Delta x/\Delta t$ being the lattice speed, and $\Delta x$ and $\Delta t$  the lattice spacing and time step, respectively; $\tau$ is the dimensionless relaxation time related to the shear viscosity $\nu$ by $\nu=c_s^2(\tau-0.5)\Delta t$ with $c_s=c/\sqrt{3}$ a model parameter; The equilibrium distribution function $f^{(eq)}$ is given by
\begin{equation}
\label{eq:feq}
f_{i}^{(eq)}=w_{i} \rho\left[1+\frac{\c_{i} \cdot \u}{c_{s}^{2}}-\frac{\left(\c_{i} \cdot \u\right)^{2}}{2 c_{s}^{4}}+\frac{\u \cdot \u}{2 c_{s}^{2}}\right],
\end{equation}
where the weights are given by $w_0=4/9$, $w_1=w_2=w_3=w_4=1/9$, and $w_5=w_6=w_7=w_8=1/36$. The density $\rho$ and velocity $\u$ are defined by
\begin{equation}
\label{eq:rhoU}
\rho=\sum_{i=0}^8{f_i},\quad \rho\u=\sum_{i=0}^8{\c_i f_i} + \dfrac{\Delta t}{2}\F,
\end{equation}
where $\F$ is the interaction force defined later; The forcing term $F_i$ in Eq. \eqref{eq:D2Q9} is defined by
\begin{equation}
\label{eq:Fi}
F_{i}=w_{i}\left[\frac{c_{i} \cdot \F}{c_{s}^{2}}+\frac{\u \F:\left(\c_{i} \c_{i}-c_{s}^{2} \I\right)}{c_{s}^{4}}\right].
\end{equation}
Note that the forcing term in \cite{ref:GuoBook} is defined as $F_i = (\c_i-\u)\cdot\F f_i^{(eq)}/c_s^2\rho$. The first and second order moments of the forcing terms of the two terms are the same, but the third-order moment are different with difference of order $O(|\u|^3)$.

The interaction force $\F$ usually depends on the free energy of the system, and can be written in either pressure form or potential form \cite{ref:Wagner2003,ref:Lee},
\begin{equation}
\F=\underbrace{\nabla(c_s^2\rho-p_0) - \kappa \nabla\nabla^2\rho}_{pressure form}
=\underbrace{\nabla(c_s^2\rho) - \rho\nabla\mu}_{potential form},
\end{equation}
where $p_0$ is the thermodynamic pressure dependent on the equation of state, $\kappa$ is a parameter relating to the surface tension, $\mu = \mu_0 - \kappa\nabla^2\rho$ is the chemical potential with $\mu_0$ being the chemical potential in bulk phase. Although the two formulations are identical mathematically, the performance of their discrete versions may be different due to discretizations errors. For instance, it was shown that the use of potential form can reduce the magnitude of SV greatly in both finite-volume and LBE methods \cite{ref:Wagner2003,ref:Lee}. In the present study we will also adopt the potential form.

It should be noted that the term $-\rho\nabla\mu$ is the physical thermodynamic driven force in the two phase system, while the term $\nabla(c_s^2\rho)$ in $\F$ is artificial which is used to cancel the ideal gas pressure induced by the collision-streaming process of the LBE. To see this more clearly, we write out the hydrodynamic equations derived from the LBE \eqref{eq:D2Q9} via the Chapman-Enskog expansion up to second order (i.e., Navier-Stokes level) \cite{ref:GuoBook},
\begin{subequations}
	\label{eq:NS}
	\begin{equation}
	\partial_t\rho+\nabla\cdot(\rho\u)=0,
	\end{equation}
	\begin{equation}
	\label{eq:D2Q9-uEq}
       \partial_t(\rho\u)+\nabla\cdot(\rho\u\u) = -\nabla(p_{id}) + \nabla\cdot\bm{\sigma} + \F,
    \end{equation}
\end{subequations}
where $p_{id}=c_s^2\rho$ is the equation of state of ideal gas, $\sigma_{\alpha\beta}=\rho\nu(\partial_\alpha u_{\beta}+\partial_\beta u_\alpha)$ is the viscous stress tensor, with $\nu=c_s^2(\tau-0.5)\Delta t$. The first term on the right hand side of Eq. \eqref{eq:D2Q9-uEq} comes from the $f^{(eq)}$ defined by Eq. \eqref{eq:feq}. Even without the forcing term, the collision-streaming procedure in the standard LBE with such $f^{(eq)}$ will also yield this term. In order to recover the desired nonideal equation of state $p_0$ or the chemical potential, it is necessary to include the additional term $\nabla(c_s^2\rho)$ in $\F$ to cancel this undesired term. However, although mathematically the term  $-\nabla p_{id}$ and the $\nabla(c_s^2\rho)$ in $\F$ can be well balanced, their discrete versions may be not due to discretization errors.

The above arguments also suggest that it is not adequate to reveal the force imbalance structure of LBE based on the derived hydrodynamic equations. This is reasonable, since the Navier-Stokes equations \eqref{eq:NS} describe just the asymptotic behavior of the LBE. Therefore, in order to obtain the structure of force imbalance of LBE, we follow the method in \cite{ref:GuoPRE2011} to analyze the discrete behavior of LBE.

Specifically, we consider a simple liquid-vapor system with a flat phase interface parallel to the $x$-direction (Fig. 1), where far from the interface the liquid and vapor phases maintain their bulk densities $\rho_l$ and $\rho_v$, respectively. For this unidirectional problem, the density, velocity, and interaction force depend only on $y$, i.e., $\rho=\rho(y)$, $\u= (u_x, u_y) =(0, v)$, and $\F=(F_x, F_y)=(0,F)$. For this system, the equilibrium state gives that $v=0$ and $\mu=\mbox{constant}$, from which the density distribution can be derived. For the LBE, the distribution functions can be categorized into three groups, namely, $A=\{f_0, f_1, f_3\}$, $B=\{f_2, f_5, f_6\}$, and $C=\{f_4, f_7, f_8\}$. In group $A$, the distribution functions are for particles moving parallel to the $x$-axis, while those in $B$ and $C$ are for particles moving upward and downward, respectively.

At steady state, the LBE \eqref{eq:D2Q9} can be written as
\begin{equation}
\label{eq:D2Q9-steady}
f_i(\x+\c_i\Delta t) - f_i (\x) = \dfrac{1}{\tau}\left(f_i-f_i^{(eq)}\right) +\Delta t \left(1-\dfrac{1}{2\tau}\right) F_i.
\end{equation}
Note that $f_i$ is independent on $x$, and so for the distribution functions in group $A$, we have
\begin{equation}
f_i=f_i^{(eq)}+\Delta t(\tau-0.5) F_i, \quad i=0, 1, 3,
\end{equation}
from which we can obtain that
\begin{equation}
f_A=f_{0}+f_{1}+f_{3}=\frac{2}{3} \rho\left(1-\frac{v^{2}}{2 c_{s}^{2}}\right)+\frac{1}{3}(1-2 \tau) \Delta t \frac{v F}{c_{s}^{2}}.
\end{equation}
Then from Eq. \eqref{eq:rhoU} we have
\begin{equation}
f_B + f_C =\rho -f_A,\quad c(f_B-f_c)=\rho v - \dfrac{\Delta t}{2} F,
\end{equation}
where $f_B=f_2+f_5+f_6$ and $f_C=f_4+f_7+f_8$. Without lose of generality, we take the lattice speed $c$ as the velocity unit (i.e., $c=1$). It is easy to obtain from the above equations that
\begin{eqnarray}
f_B=\dfrac{\rho}{6}+\dfrac{\rho v}{2}+\frac{\rho v^{2}}{2}-\dfrac{\Delta t}{4} F+(\tau-0.5) \Delta t F v, \\
f_C=\dfrac{\rho}{6}-\dfrac{\rho v}{2}+\frac{\rho v^{2}}{2}+\dfrac{\Delta t}{4} F+(\tau-0.5) \Delta t F v.
\end{eqnarray}
With these results, we multiply Eq. \eqref{eq:D2Q9-steady} by $c_{iy}$ and take summation over $i$ to obtain the following discrete momentum balance equation,
\begin{equation}
f_{B,j+1}-f_{C, j-1}=\rho_j v_j+\frac{\Delta t}{2} F_j,
\end{equation}
or
\begin{equation}
\label{eq:momentumBalance}
-\dfrac{c_{s}^{2}\left(\rho_{j+1}-\rho_{j-1}\right)}{2 \Delta x}+\dfrac{F_{j+1}+2 F_{j}+F_{j-1}}{4}=-\Delta x R(v_j),
\end{equation}
where $j$ is the index of the lattice nodes in $y$-direction, and
\begin{equation}
\label{eq:Rv}
R(v_j)=\dfrac{\rho_{j+1} v_{j+1}-2 \rho_{j} v_{j}+\rho_{j-1} v_{j-1}}{2 \Delta x^{2}}+(2 \tau-1) \dfrac{\left(F_{j+1} v_{j+1}-F_{j-1} v_{j-1}\right)}{2 \Delta x}+\dfrac{\rho_{j+1} v_{j+1}^{2}-\rho_{j-1} v_{j-1}^{2}}{2 \Delta x}.
\end{equation}
Note that Eq. \eqref{eq:momentumBalance} represents the total force balance at discrete level, which can be viewed as a finite-difference scheme of the continuous momentum equation \eqref{eq:D2Q9-uEq}. This discrete balance equation has been obtained in \cite{ref:GuoBook} except for a slight difference in $R(v)$ at the order of $O(v^3)$. However, in that work the detailed discretization errors were not identified.

\begin{figure}
\includegraphics[width=0.48\textwidth]{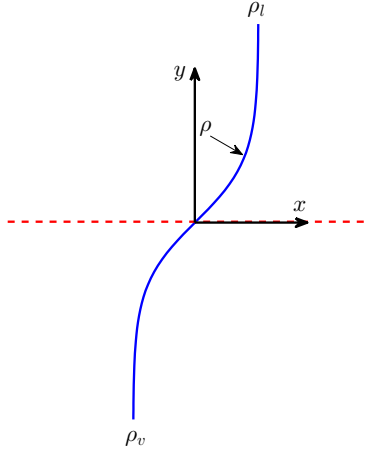}
\caption{(Color online) Sketch of the flat interface problem.} \label{fig:FlatInterface}.
\end{figure}

If the LBE is well-balanced at equilibrium, the equilibrium state ${\cal S}$ should be the solution of Eq. \eqref{eq:momentumBalance}. Unfortunately, this does not hold in general.
We will demonstrate this point using the method of proof by contradiction. If ${\cal S}$ is the solution of \eqref{eq:momentumBalance}, then we have $R(v_j)=0$ and $F_j =\partial_y(c_s^2\rho)_j =c_s^2\delta_h \rho_j$, where $\delta_h$ is certain discrete gradient operator, which can be expressed generally as
$$
\delta_h\rho_j=\dfrac{1}{\Delta x}\sum_{k=-r}^{s}{a_{j+k}\rho_{j+k}},
$$
where $r$ and $s$ are two positive integers, and $a_i$ are nonzero constant coefficients. Equation \eqref{eq:momentumBalance} then leads to
\begin{equation}
\label{eq:error}
\dfrac{\rho_{j+1}-\rho_{j-1}}{2 \Delta x}+\delta_h^F\rho_j = 0,
\end{equation}
where
\begin{equation}
 \delta_h^F\equiv\dfrac{\delta_h\rho_{j+1}+2 \delta_h\rho_{j}+\delta_h\rho_{j-1}}{4}.
\end{equation}

The first and second terms of Eq. \eqref{eq:error} are both discretizations of $\partial_y(c_s^2\rho)$.  However, their stencils and truncation errors are different generally. For instance, if the widely used isotropic central scheme (ICS) \cite{ref:GuoPRE2011} is employed to discretize the gradients in $\F$, we have
\begin{equation}
\delta_h \rho_j = \dfrac{\rho_{j+1}-\rho_{j-1}}{2\Delta x},
\end{equation}
and the second term of Eq. \eqref{eq:error} can be written as
\begin{equation}
\label{eq:drho-F}
\delta_h^F \rho_j = \dfrac{\rho_{j+2}+2(\rho_{j+1}-\rho_{j-1})-\rho_{j-2}}{8\Delta x}\approx \partial_y\rho +\dfrac{5\Delta x^2}{12}\partial_y^3\rho.
\end{equation}
On the other hand, the first term coming from the evolution of LBE can be written as
\begin{equation}
\label{eq:drho-LB}
-\delta_h^{LB}\rho_j = -\dfrac{\rho_{j+1}-\rho_{j-1}}{2 \Delta x} \approx -\partial_y\rho -\dfrac{\Delta x^2}{6}\partial_y^3\rho.
\end{equation}
Obviously, the truncation errors of $\delta_h^{LB}\rho_j$ (from LBE) and $\delta_h^F\rho_j$ (from interaction force) cannot be canceled with each other, except as $\rho$ is a second polynomial of $y$. However, if $\rho(y)=a_2 y^2+a_1 y+a_0$ with $a_i$ being some constants, the boundary conditions $\rho(-\infty) =\rho_v$ and $\rho(\infty)=\rho_l$ suggest that $a_1=a_2=0$, meaning that the density is a constant. This is clearly impossible for a two phase system with different bulk densities.

Other discretization schemes of $\delta_h\rho_j$ other than ICS will lead to different formulation of $\delta_h^F\rho_j$, but the stencil of $\delta_h^F\rho_j$ is usually larger than that of $\delta_h^{LB}\rho_j$, and their truncation errors cannot be canceled with each other. Therefore, in general the force balance equation \eqref{eq:error} does not hold, and we complete the proof that the equilibrium state ${\cal S}$ cannot be the solution of the discrete momentum balance equation \eqref{eq:momentumBalance} of the standard LBE.

A byproduct of the above proof is the structure of the net force due to the discretization errors in the standard LBE, namely
\begin{equation}
\label{eq:netForce}
\Delta F = -\delta_h^{LB}\rho_j + \delta_h^F\rho_j,
\end{equation}
which depends on the stencil of $\delta_h$. Particularly, the net force with the ICS discretization is
\begin{equation}
\Delta F \approx \dfrac{\Delta x^2}{12}\partial_y^3\rho.
\end{equation}
suggesting that deviation of the LBE solution from the equilibrium state ${\cal S}$ depends on mesh size and density difference. It is noted that in some studies high-order discretizations for the interaction $\F$ has been employed (e.g., \cite{ref:Shan,ref:FromPRE2020}). Although the high-order schemes can increase the isotropy of discretization and reduce the magnitude of SV, but the improvement is limit. This can also be explained from the structure of net force \eqref{eq:netForce}: the force balance cannot be reached by improving the accuracy of $\delta_h^F$ solely.

\section{A well-balanced LBE}
Based on the above analysis, we now present a well-balanced LBE that can capture the equilibrium state.
Since the different truncation errors in $\delta_h^{LB}\rho$ and $\delta_h^{F}\rho$ are the main barrier for the standard LBE to reach the equilibrium state, it is necessary to remove their effects in order to capture the correct thermodynamic properties of the two-phase system. As indicated in Sec. II, $\delta_h^{LB}\rho$ comes from the evolution of LBE with the equilibrium distribution function defined by Eq. \eqref{eq:feq}, and $\delta_h^F\rho$ in $F$ is used to cancel it. Therefore, if $\delta_h^{LB}\rho$ does not appear, $\delta_h^F\rho$ is not required yet, and the force imbalance in the standard LBE may be avoided.

A more careful examination shows that $\delta_h^{LB}\rho$ is induced by the first term ($\rho$) in $f^{(eq)}$ given by Eq. \eqref{eq:feq}. Therefore, to avoid the appearing of  $\delta_h^{LB}\rho$, we redefine the equilibrium distribution function as
\begin{equation}
\label{eq:mFeq}
f_i^{(eq)}=\left\{
\begin{array}{ll}
\rho-(1-w_0)\rho_0-w_0 \rho \dfrac{\u \cdot \u}{2 c_s^2}, & i=0 \\
w_i\left\{\rho_0 + \rho\left[\dfrac{\c_i \cdot \u}{c_s^2}+\dfrac{\left(\c_i \cdot \u\right)^2}{2 c_s^4}-\dfrac{\u \cdot \u}{2 c_s^2}\right]\right\}, & i \neq 0
\end{array}
\right.
\end{equation}
where $\rho_0$ is a numerical constant. The choice of $\rho_0$ may influence the stability of the scheme, but does not influence the theoretical results. In what follows, we simply set $\rho_0=0$.
Accordingly, the interaction force is defined now by
\begin{equation}
\label{eq:mF}
\F=-\rho\nabla\mu,
\end{equation}
and the forcing term $F_i$ is modified as
\begin{equation}
\label{eq:mFi}
F_{i}=w_{i}\left[\frac{c_{i} \cdot \F}{c_{s}^{2}}+\frac{\u (\F+c_s^2\nabla\rho):\left(\c_{i} \c_{i}-c_{s}^{2} \I\right)}{c_{s}^{4}} + \dfrac{1}{2}\left(\dfrac{c_i^2}{c_s^2}-D\right)(\u\cdot\nabla\rho)\right],
\end{equation}
where $D$ is the dimension and $D=2$ for the D2Q9 model. It is easy to verify the following moments of $f^{(eq)}$ and $F_i$ defined above,
\begin{equation}
\sum_i{f_i^{(eq)}} =\rho,\quad \sum_i{\c_i f_i^{(eq)}} =\rho\u,\quad \sum_i{\c_i\c_i f_i^{(eq)}} =\rho\u\u,\quad \sum_i{\c_i\c_i\c_i f_i^{(eq)}} =c_s^2\rho[\u\bm{I}],
\end{equation}
\begin{equation}
\sum_i F_i =0, \quad \sum_i \c_i F_i=\F, \quad \sum_i \c_i\c_i F_i = [\u\F] + c_s^2\left([\u\nabla\rho]+(\u\cdot\nabla\rho) \bm{I}\right),
\end{equation}
where $[\u\bm{I}]_{\alpha\beta\gamma} = u_\alpha\delta_{\beta\gamma} + u_\beta\delta_{\alpha\gamma} +u_\gamma\delta_{\alpha\beta}$, $[\u\F]_{\alpha\beta}=u_\alpha F_\beta + u_\beta F_\alpha$, and $[\u\nabla\rho]_{\alpha\beta}=u_\alpha \partial_\beta\rho + u_\beta \partial_\alpha\rho$. With these moments, it can be derived the hydrodynamic equations at the Navier-Stokes level via the Chapman-Enskog expansion,
\begin{subequations}
	\label{eq:mNS}
	\begin{equation}
	\partial_t\rho+\nabla\cdot(\rho\u)=0,
	\end{equation}
	\begin{equation}
	\label{eq:mD2Q9-uEq}
	\partial_t(\rho\u)+\nabla\cdot(\rho\u\u) = -\rho\nabla\mu + \nabla\cdot\bm{\sigma}',
	\end{equation}
\end{subequations}
where $\sigma'_{\alpha\beta}=\rho\nu[\partial_\alpha u_\beta + \partial_\beta u_\alpha + (\nabla\cdot\u)\delta_{\alpha\beta}]$, which is slightly different from $\sigma_{\alpha\beta}$ of the standard LBE in the bulk viscosity.

It is clear that the new LBE with the modified $f^{(eq)}$, $\F$, and $F_i$ can yield the desired asymptotic hydrodynamic equations, which can describe equilibrium state of a two-phase system. To see whether the LBE can capture the state at the discrete level, we again consider the flat interface problem as considered in Sec. II. Following the same procedure, we can obtain that
\begin{eqnarray}
f_A=\rho(1-v^2) + (1-2\tau) \Delta t Fv - (1-2\tau)\Delta t \left(F + 0.5\delta_h\rho\right) \\
f_B=\dfrac{\rho v}{2}+\frac{\rho v^{2}}{2}-\dfrac{\Delta t}{4} F+(\tau-0.5) \Delta t (F+0.5\delta_h\rho) v, \\
f_C=-\dfrac{\rho v}{2}+\frac{\rho v^{2}}{2}+\dfrac{\Delta t}{4} F+(\tau-0.5) \Delta t (F+0.5\delta_h\rho) v,
\end{eqnarray}
and the momentum balance equation is
\begin{equation}
\label{eq:mBalance}
\dfrac{\rho_{j+1}\delta_h\mu_{j+1}+2\rho_j \delta_h\mu_{j}+\rho_{j-1}\delta_h\mu_{j-1}}{4}=\Delta x R'(v_j),
\end{equation}
where
\begin{equation}
\begin{split}
\label{eq:mRv}
R'(v_j)&=\dfrac{\rho_{j+1} v_{j+1}-2 \rho_{j} v_{j}+\rho_{j-1} v_{j-1}}{2 \Delta x^{2}}+(2 \tau-1) \dfrac{\left(F_{j+1} v_{j+1}-F_{j-1} v_{j-1}\right)}{2 \Delta x}+\dfrac{\rho_{j+1} v_{j+1}^{2}-\rho_{j-1} v_{j-1}^{2}}{2 \Delta x} \\
&\qquad + (\tau-0.5) \dfrac{(\delta_h\rho_{j+1}) v_{j+1}-(\delta_h\rho_{j-1}) v_{j-1}}{2 \Delta x}.
\end{split}
\end{equation}

It is easy to verify that $v_j=0$ and $\mu_j=\mbox{constant}$ is a solution of Eq. \eqref{eq:mBalance}, suggesting that the new LBE can reach the equilibrium state ${\cal S}$ at discrete level. Therefore, the new LBE \eqref{eq:D2Q9} together with Eqs. \eqref{eq:mFeq}, \eqref{eq:mF}, and \eqref{eq:mFi} can be well balanced at equilibrium, and we will term it as {\em well-balanced} LBE (WB-LBE).

\section{Numerical results}
We now test the well-balanced properties of the WB-LBE by simulating two cases. The free-energy density functional of the system is chosen to be
\begin{equation}
{\cal F}=\int{ \left[\psi_0(\rho) + \dfrac{1}{2}\kappa|\nabla\rho|^2 \right] dV},
\end{equation}
where the bulk energy $\psi_0$ takes the double-well form,
\begin{equation}
\label{eq:doubleWell}
\psi_0 = \beta (\rho-\rho_l^0)^2(\rho-\rho_v^0)^2,
\end{equation}
where $\beta$ is a constant, $\rho_l^0$ and $\rho_v^0$ are the densities of liquid and vapor phases at saturation, respectively. The equation of state and bulk chemical potential are given respectively by
\begin{equation}
\label{eq:EOS}
p_0=\dfrac{d\psi_0}{d\rho}-\psi_0, \quad \mu_0 = \dfrac{d\psi_0}{d\rho}.
\end{equation}

\begin{figure}
\includegraphics[width=0.3\textwidth]{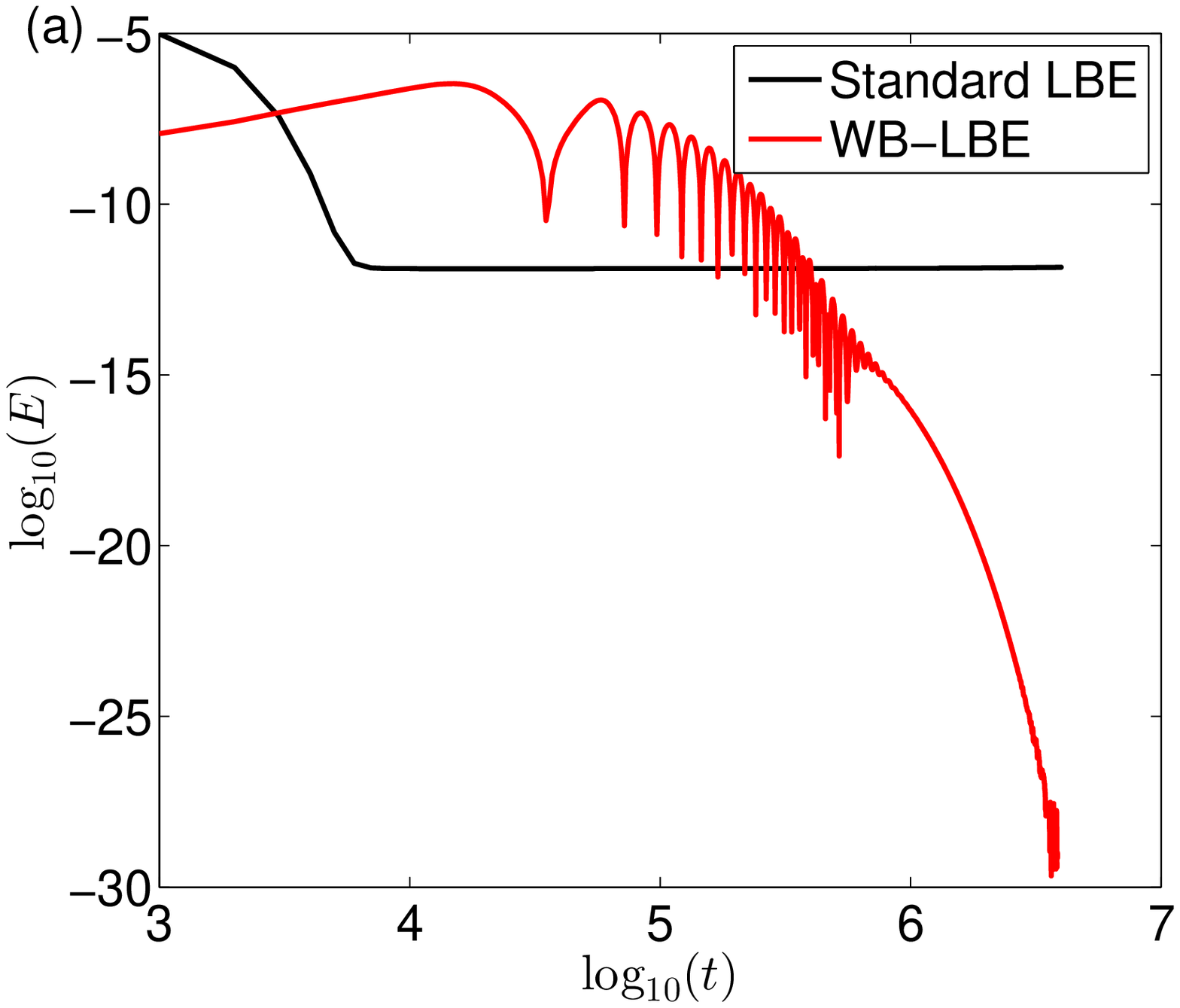}
\includegraphics[width=0.3\textwidth]{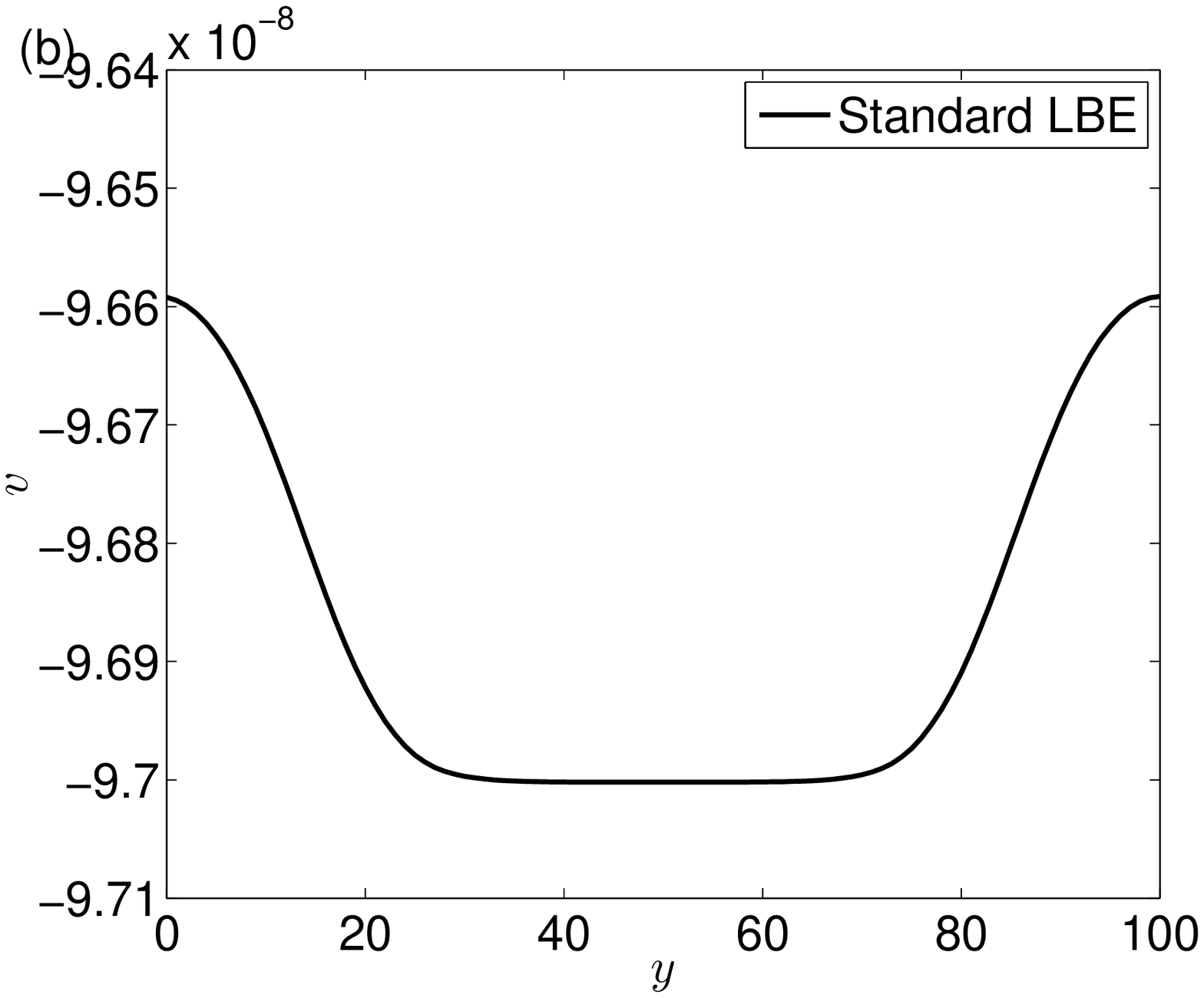}
\includegraphics[width=0.3\textwidth]{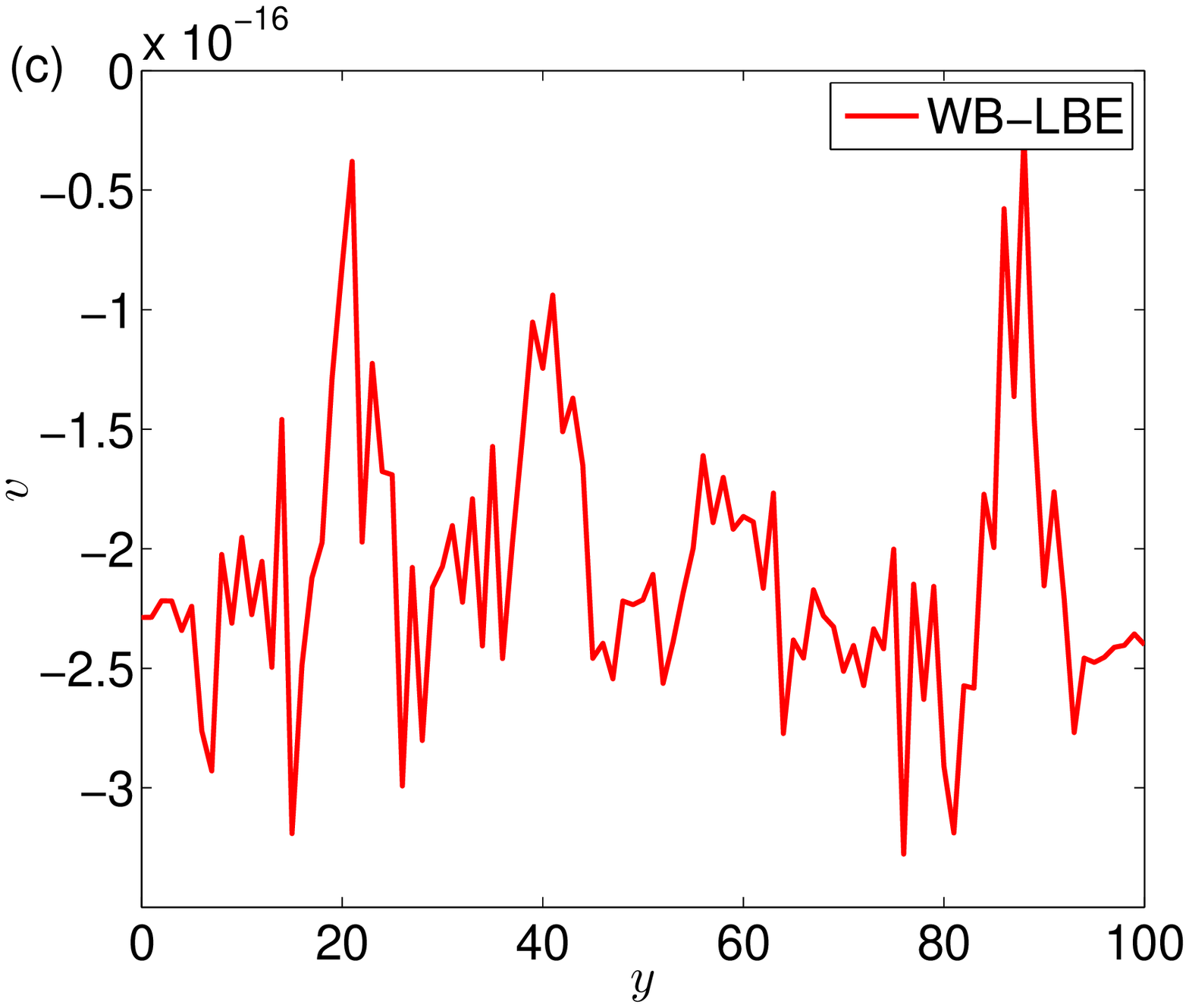}
\caption{(Color online) Results of the flat interface problem. (a) Time history of the kinetic energy; (b) Velocity of the standard LBE at steady state; (c) Velocity of the WB-LBE at steady state.} \label{fig:E-1D}
\end{figure}

\subsection{1D flat interface}
 The first test case is a flat interface problem, where the computation domain is covered by a $N_x\times N_y =21\times 101$ lattice and the liquid slab is put in the central region ($25\le y \le 75$) and the rest is filled with the vapor phase. Periodic boundary conditions are employed at the four boundaries.
 In the simulations, we choose $\rho_l^0=1.0$, $\rho_v^0=0.2$, $\beta=0.01$, and $\kappa=0.0128$. For this plane interface problem, the density at equilibrium follows a hyperbolic tangent profile,
\begin{equation}
\label{eq:Tanh}
\rho_0(y)=\rho_v^0 +\dfrac{\rho_l^0-\rho_v^0}{2}\left[ \tanh(2(y-y_1)/D) - \tanh(2(y-y_2)/D) \right],
\end{equation}
where $y_1$ and $y_2$ are the positions of the lower and upper interfaces, $D=(\rho_l-\rho_v)\sqrt{8\kappa/\beta}$ characterizes the interface thickness. The relaxation time $\tau$ is set to be 0.85 in all simulations. Initially, the density is specified with the theoretical solution \eqref{eq:Tanh} with a small random perturbation, $\rho(y)=(1+\epsilon r)\rho_0(y)$, with $\epsilon=0.01$ and $r$ a random number sampled between $-1$ and 1 from a uniform distribution.

\begin{figure}
\includegraphics[width=0.45\textwidth]{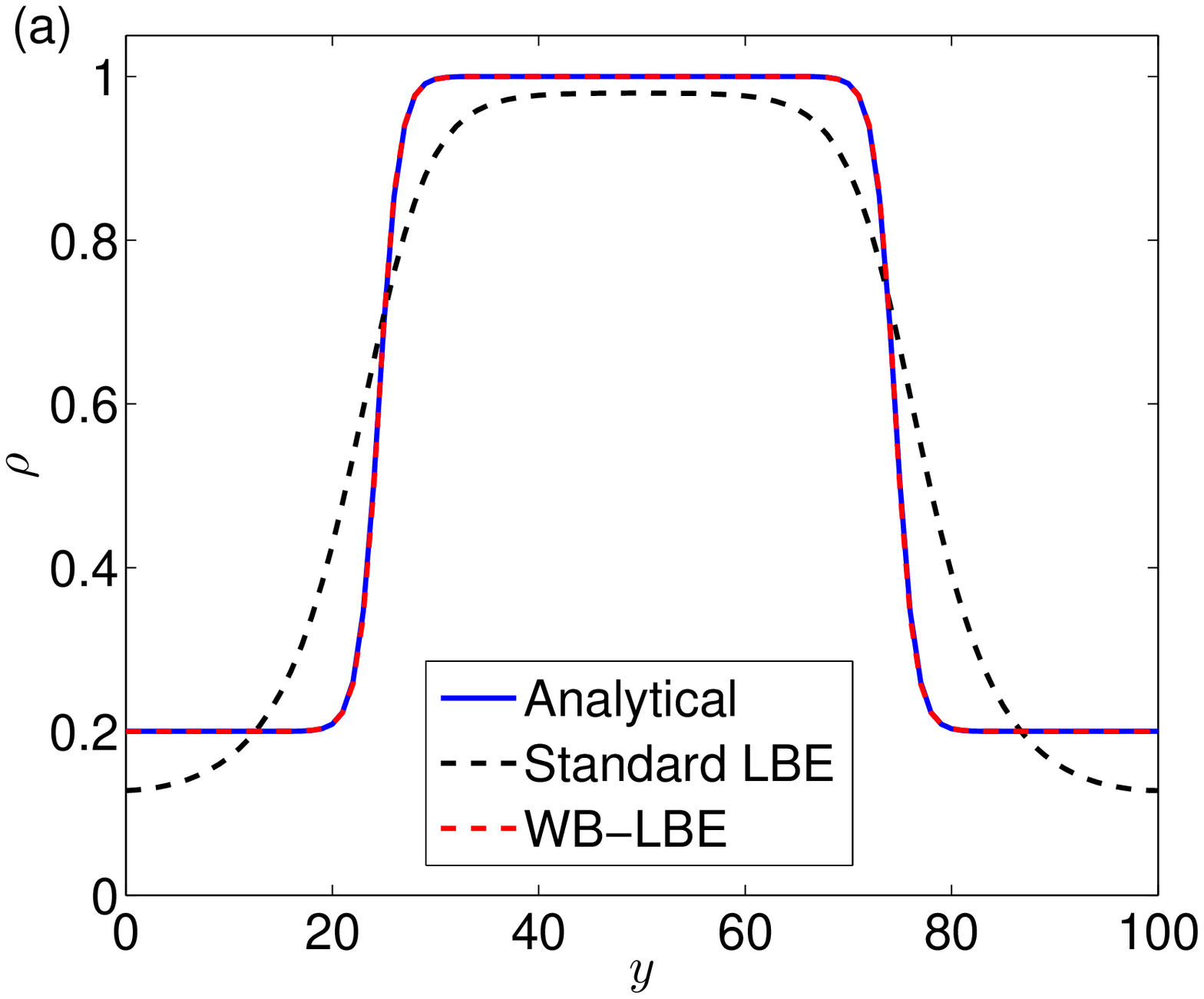}
\includegraphics[width=0.45\textwidth]{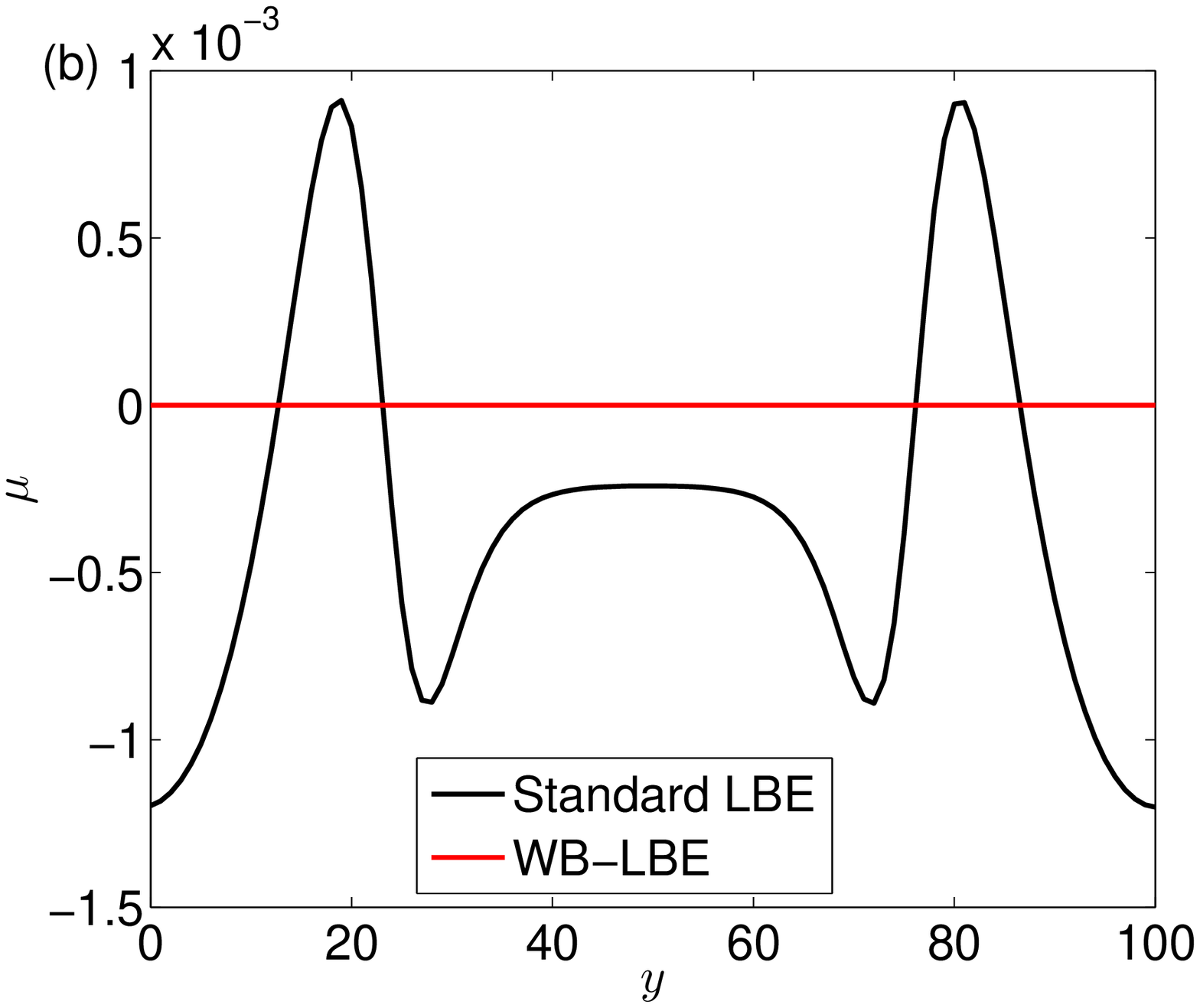}
\caption{(Color online) Density (a) and chemical potential (b) at the steady state of the flat interface problem.} \label{fig:density-1D}
\end{figure}

The time history of the system kinetic energy, $E(t)=\tfrac{1}{2}\int{\rho|\u|^2 d\x}$, are shown in Fig. \ref{fig:E-1D}(a). For the standard LBE, the kinetic energy reaches to a nearly constant value at the order of $10^{-12}$. On the other, the kinetic energy from the WB-LBE deceases to about $10^{-30}$. The velocity profiles of both LBE models are shown in Fig.  \ref{fig:E-1D}(b) and  \ref{fig:E-1D}(c). It can be seen that the standard LBE predicts a velocity field of order $10^{-8}$, which clearly demonstrate the existence of SV; while for the WB-LBE, the SV is removed up to the machine accuracy. The density and chemical potential distributions are shown in Fig. \ref{fig:density-1D}. The density profile predicted by the WB-LBE is nearly indistinguishable from the theoretical one, while that predicted by the standard LBE shows clearly deviations, particularly in the vapor phase. The chemical potential predicted by the WB-LBE is nearly constant, while that predicted by the standard LBE varies at the order of $10^{-3}$. The above results clearly show that the present WB-LBE can reach the equilibrium state for this flat interface problem, while the standard one fails, which is consistent with the theoretical analysis.

\subsection{2D droplet}
\begin{figure}
\includegraphics[width=0.3\textwidth]{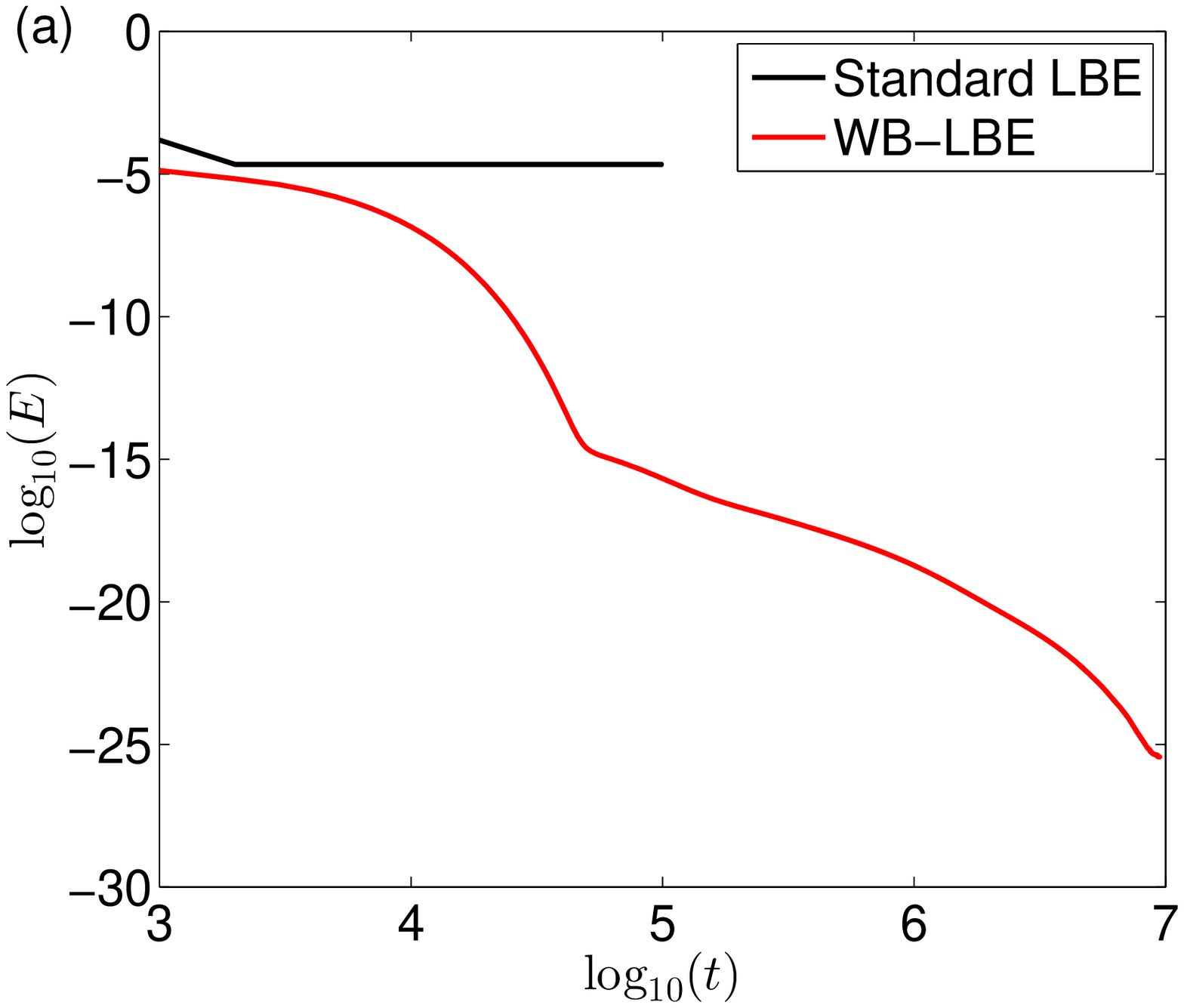}
\includegraphics[width=0.3\textwidth]{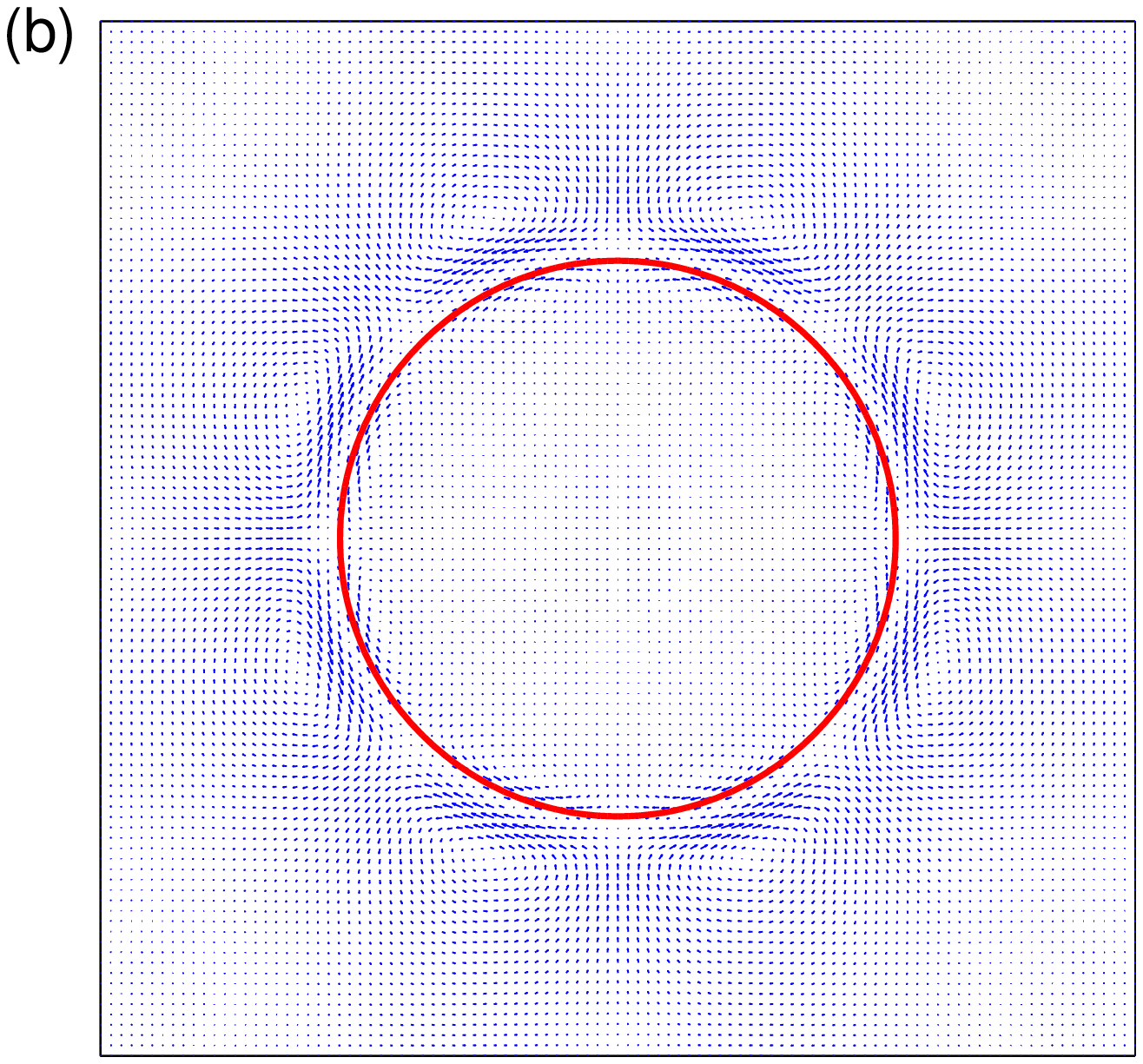}
\includegraphics[width=0.3\textwidth]{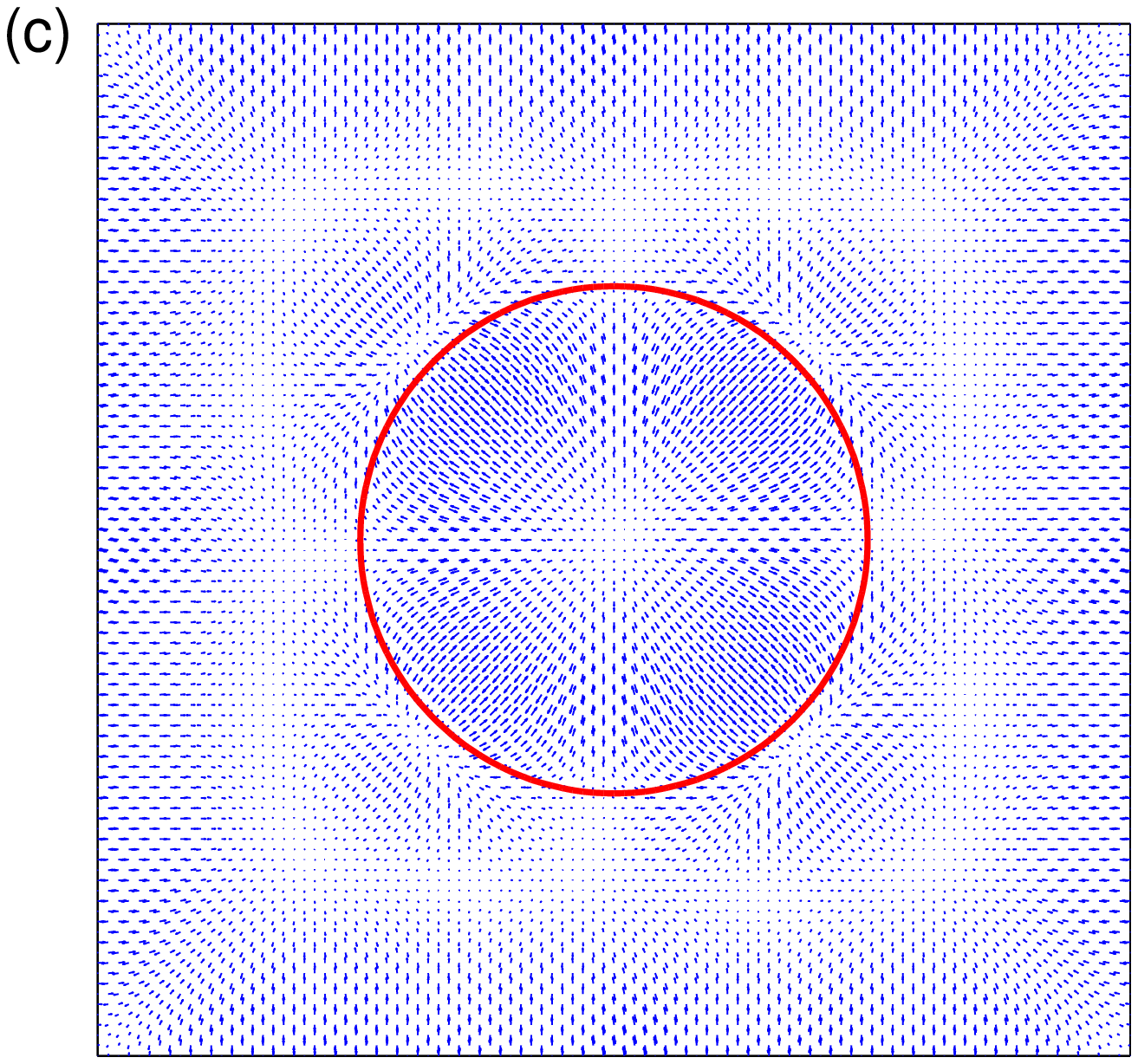}
\caption{(Color online) (a) Time history of the kinetic energy; (b) Velocity field of the standard LBE. The maximum magnitude is $4.78\times 10^{-4}$; (c) Velocity field of the standard LBE. The maximum magnitude is $8.63\times 10^{-15}$. The red circle is the density contour plot at $(\rho_l+\rho_v)/2$. } \label{fig:E-2D}
\end{figure}

The second test case is a 2D droplet immersed in its vapor in a square domain of size $100\times 100$. The radius of the droplet is 25, and is initialized at the center of domain. Periodic boundary conditions are applied to all boundaries, and the simulation parameters are the same as used in the 1D flat interface problem. Figure \ref{fig:E-2D} presents the evolutions of the system kinetic energy, and the velocity fields predicted by the standard LBE and WB-LBE at steady state. It is clear that the WB-LBE can produce negligible SV, while the SV in standard LBE is quite large.

\begin{figure}
\includegraphics[width=0.3\textwidth]{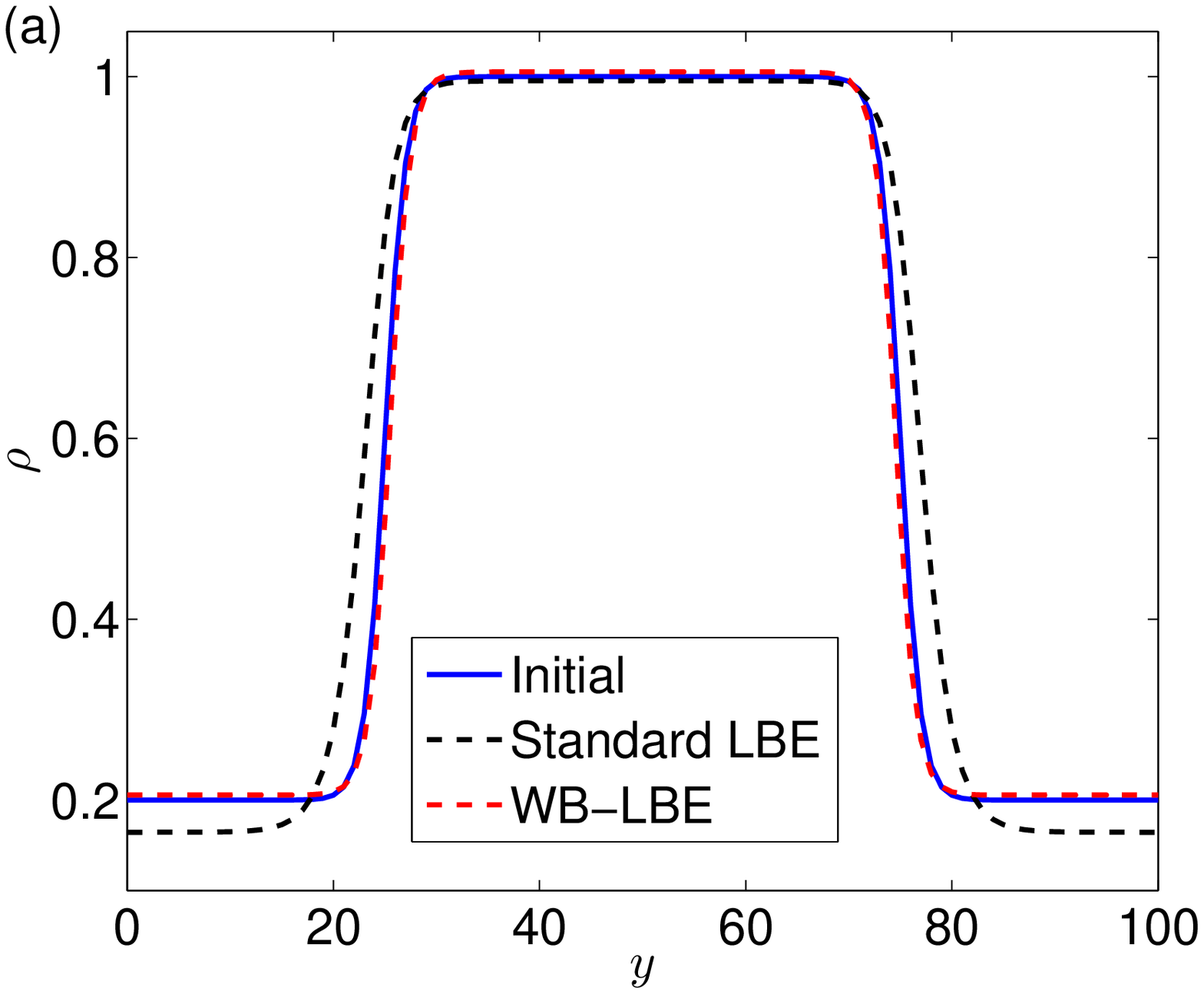}
\includegraphics[width=0.3\textwidth]{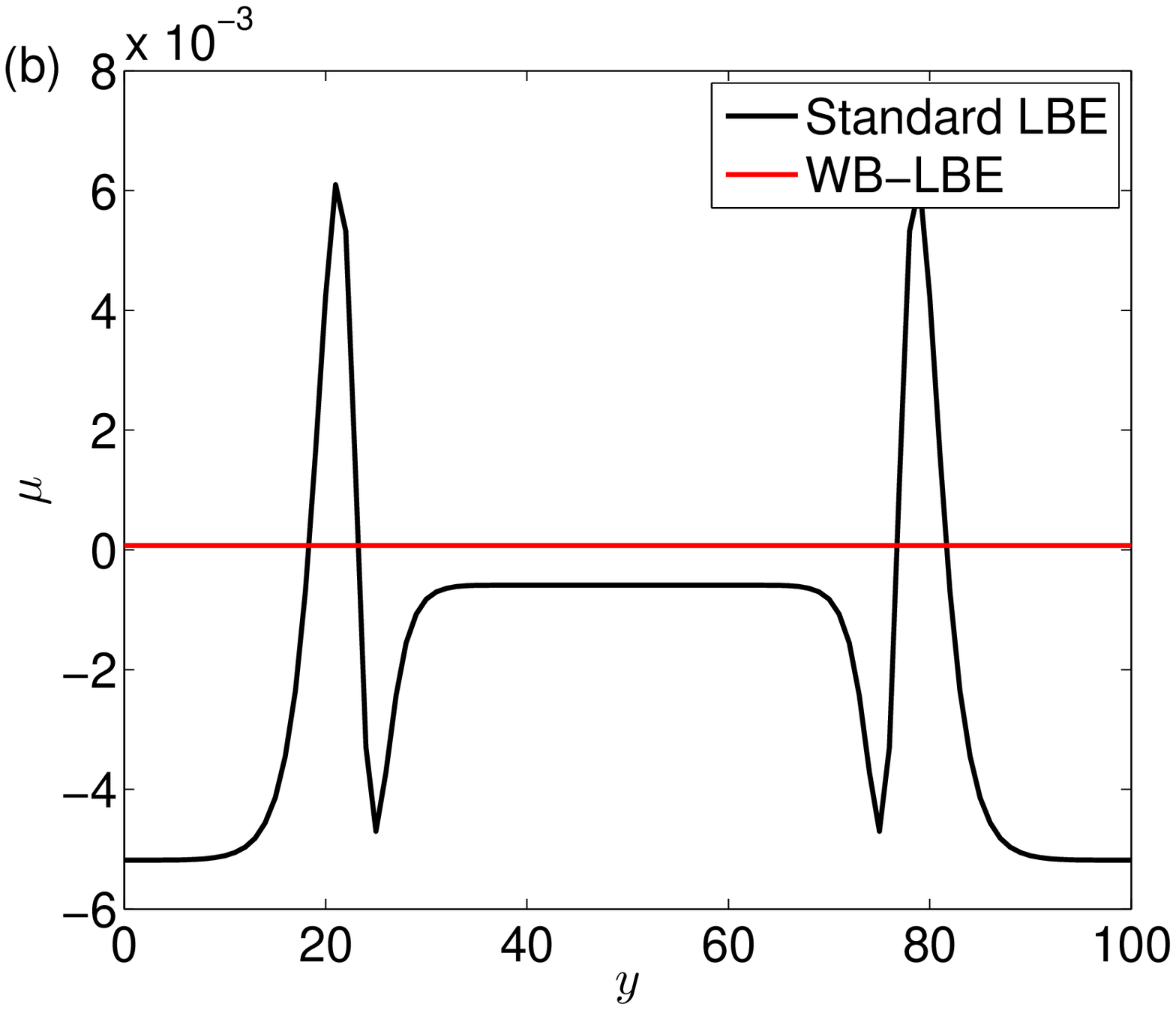}
\includegraphics[width=0.3\textwidth]{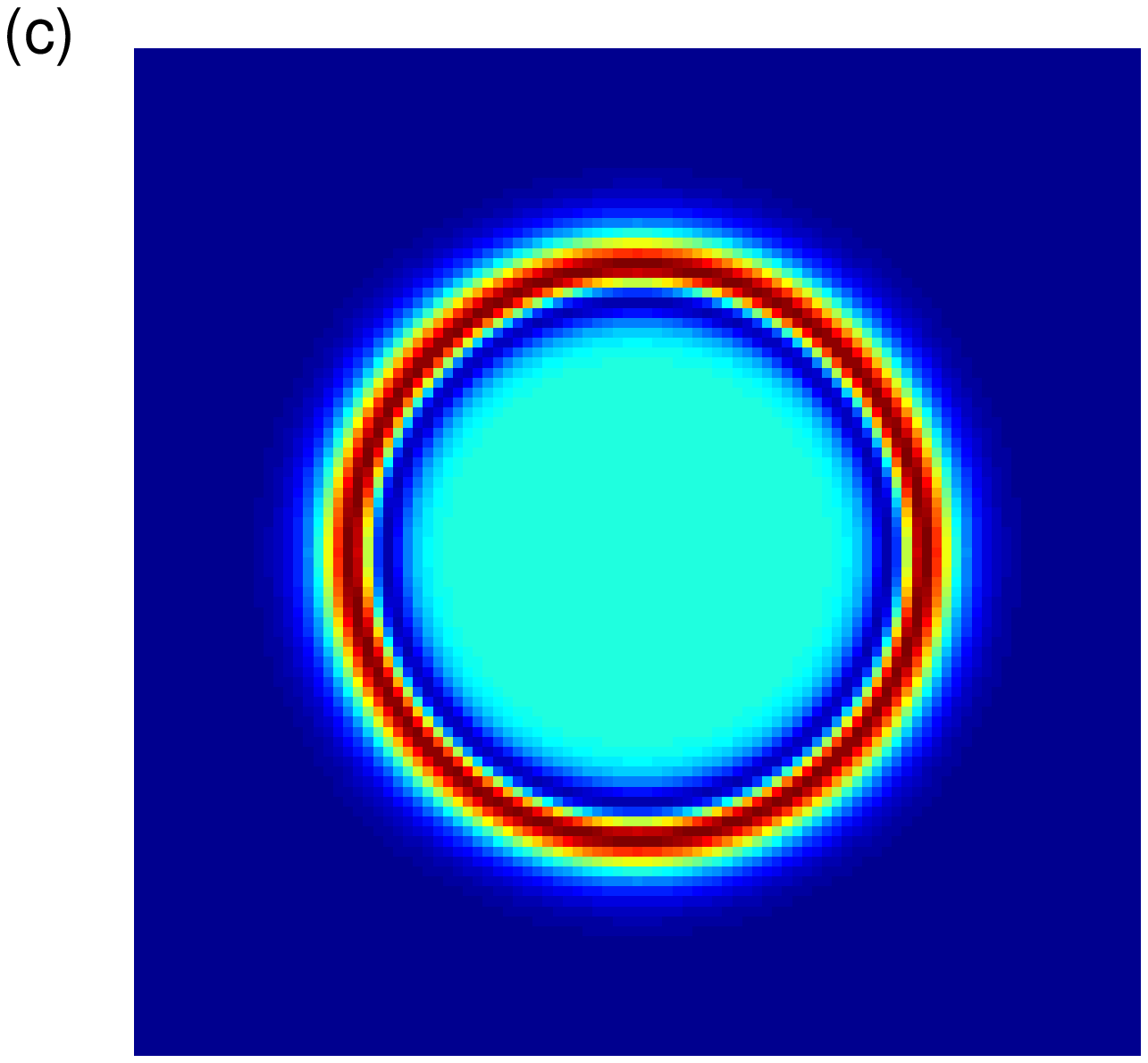}
\caption{(Color online) Results of the 2D droplet problem at steady state. Density (a) and chemical potential (b) profiles across the droplet center; (c) The distribution of chemical potential of the standard LBE.} \label{fig:density-2D}
\end{figure}

The density and chemical potential structure are shown in Fig. \ref{fig:density-2D}. It can be seen that from Fig. \ref{fig:density-2D}(a) that the density profile from the WB-LBE agree well with the initial profile, but the bulk liquid and vapor densities are slightly larger than the corresponding saturation values. The increase is due to the curvature of the droplet, which can be obtained from the bulk pressure values determined by the Laplace's law\cite{ref:Lee,ref:LouPRE}. However, the bulk density of the vapor phase predicted by the standard LBE deviates from the theoretical ones greatly, and the interface thickness becomes wider than the theoretical values. Figure \ref{fig:density-2D}(b) suggests that the WB-LBE predicted a nearly constant chemical potential, while the standard LBE gives a nonuniform one with variation of order $10^{-3}$, particularly near the interface as shown in \ref{fig:density-2D}(c).

%\begin{subequations}
%\begin{equation}
%p_l=p_l^0+\frac{\rho_l^0}{\rho_l^0-\rho_v^0} \dfrac{\sigma}{R}, \\
%\end{equation}
%\begin{equation}
%p_v=p_v^0+\frac{\rho_v^0}{\rho_l^0-\rho_v^0} \dfrac{\sigma}{R},
%\end{equation}
%\end{subequations}
%where $p_l^0$ and $p_v^0$ are the pressures evaluated at the saturation densities, respectively, and $\sigma$ is the surface tension. For the double-well free energy, $\sigma=\tfrac{1}{6}(\rho_l^0-\rho_v^0)^3\sqrt{2\beta\kappa}$. After obtaining $p_l$ and $p_v$, the bulk densities $\rho_l$ and $\rho_v$ can be obtained from the equation of state given by Eq. \eqref{eq:EOS}. With the parameters used in the test case, we can obtain that $\rho_l=$ and $\rho_v=$

\section{Summary}
The capability of capturing physical equilibrium state of a two-phase fluid system is a fundamental requirement of LBE. In this work a rigorous analysis is made to track the origin of force imbalance in the standard LBE at discrete level. The detailed structure of the net force is derived. It is shown that the net force comes from the discretization errors of two parts: the gradient of $-p_{id}$ coming from the evolution of LBE, and the term $\nabla(\c_s^2\rho)$ in the interaction force. Generally, the stencils of these two discretizations are different, and the truncation errors cannot be canceled with each other. This net force drives the system to deviate the physical equilibrium state, producing the spurious velocity and non-constant chemical potential or unphysical density profiles.

Motivated by the theoretical analysis, a well-balanced LBE is proposed which has the same algorithm structure as the standard LBE but avoids the incompatible discretizations. Mathematical analysis shows that the LBE can achieve the force balance condition (well-balanced), and can capture the equilibrium state at discrete level. The numerical tests of a flat interface and a droplet show that the spurious velocity can be removed, and consistent interface profile and bulk densities can be captured, confirming the well-balanced properties of the model.

It is noted that the strategy adopted in the present well-balanced LBE can also be employed in other LBE models, such as pseudo-potential and phase-field models.

%\section*{Acknowledgement}
\acknowledgements
This work was supported by the National Science Foundation of China (Grant No. 51836003).


\begin{thebibliography}{50}
\bibitem{ref:SucciBook} S. Succi, {\it The Lattice Boltzmann Equation for Fluid Dynamics and Beyond} (Oxford University Press, Oxford, 2001).
\bibitem{ref:GuoBook}  Z. L. Guo and C. Shu, {\it Lattice Boltzmann Method and its Applications in Engineering} (World Scientific Publishing, Singapore,2013).
\bibitem{ref:Cristea} A. Cristea and V. Sofonea, Int. J. Mod. Phys. C {\bf 14}, 1251 (2003).
\bibitem{ref:Wagner2003} A. J. Wagner, Int. J. Mod. Phys. B {\bf 17}, 193 (2003).
\bibitem{ref:Lee} T. Lee and P. F. Fischer, Phys. Rev. E {\bf 74}, 046709 (2006).
\bibitem{ref:Shan} X. Shan, Phys. Rev. E {\bf 73}, 047701 (2006).
\bibitem{ref:LeeRev} K. Connington and T.H. Lee, J. Mech. Sci. Tech. {\bf 26},  3857 (2012).
\bibitem{ref:Gong2019} J. Gong, N. Oshima, Y. Tabe, Comput. Math. Appl. {\bf 78}, 1166 (2019).


%---------
\bibitem{ref:Zheng2005} H. W. Zheng, C. Shu, and Y. T. Chew, Phys. Rev. E {\bf 72}, 056705 (2005).
\bibitem{ref:Wagner2006}  A. J. Wagner, Phys. Rev. E {\bf 74}, 056703 (2006).
\bibitem{ref:Kupershtokn2009}  A. L. Kupershtokn, D. A. Medvedev, and D. I. Karpov, Comput. Math. Appl. {\bf 58}, 965 (2009).
\bibitem{ref:HuangHBPRE2011} H.B. Huang, M. Krafczyk, and X.Y. Lu, Phys. Rev. E {\bf 84}, 046710 (2011).
\bibitem{ref:Sun2012} K. Sun, T. Wang, M. Jia, G. Xiao, Physica. A {\bf 391}, 3895 (2012).
\bibitem{ref:Siebert2014} D. N. Siebert, P. C. Philippi and K. K. Mattila, Phys. Rev. E {\bf 90}, 053310 (2014).
\bibitem{ref:LuoKH2015} D. Lycett-Brown and K.H. Luo, Phys. Rev. E {\bf 91}, 023305 (2015).
\bibitem{ref:ChenBX2015} S. Khajepor, J. Wen, and B.X. Chen, Phys. Rev. E {\bf 91}, 023301 (2015).
\bibitem{ref:ZhengLin2017} Q. Zhai, L. Zheng, and S. Zheng, Phys. Rev. E {\bf 95}, 023313 (2017).
\bibitem{ref:HuangPRE2019} R.Z. Huang, H.Y. Wu, and N.A. Adams, Phys. Rev. E {\bf 99}, 023303 (2019).
\bibitem{ref:Khar2019} S.F. Kharmiani, H. Niazmand, M. Passandideh-Fard, J. Stat. Phys. {\bf 175}, 47-70 (2019).
\bibitem{ref:WenPRE2020} B.H. Wen, L. Zhao, W. Qiu, Y. Ye, and X.W. Shan, Phys. Rev. E {\bf 102}, 013303 (2020).
\bibitem{ref:WagnerPRE2020} L.E. Czelusniak, V.P. Mapelli, M.S. Guzella, L. Cabezas-Gomez, and A.J. Wagner, Phys. Rev. E {\bf 102}, 033307 (2020).

%ref:Zheng2005,ref:Wagner2006,ref:Kupershtokn2009,ref:HuangHBPRE2011,ref:Sun2012,ref:Siebert2014,ref:LuoKH2015,ref:ChenBX2015,ref:ZhengLin2017,ref:HuangPRE2019,ref:Khar2019,ref:WagnerPRE2020,ref:WenPRE2020


\bibitem{ref:FromPRE2020} C.S. From, E. Sauret, S.A. Galindo-Torres, and Y. T. Gu, Phys. Rev. E {\bf 101}, 033303 (2020).

\bibitem{ref:GuoPRE2011} Z.L. Guo, C.G. Zheng, and B.C. Shi, Phys. Rev. E {\bf 83}, 036707 (2011).

\bibitem{ref:LouPRE} Q. Lou and Z.L. Guo, Phys. Rev. E {\bf 91}, 013302 (2015).




\end{thebibliography}
\end{document}